%   FSU-SCRI-92-40
%   submitted to Nucl. Phys. B
%   should be mailed out on the SCRI physics list
%
%   cover page appended following \bye
%
\input midinsert
\input preprint92.format
\null
\vskip.2in
\rightline{Preprint FSU-SCRI-92-40}
\vskip1in
\cbox{\bf Monte Carlo calculation of the surface free energy
for the 2d 7-state Potts Model,
and an estimate for 4d SU(3) gauge theory*}
\vfootnote*{\tenpoint This research project was partially funded by the
Department of Energy under Contracts DE-FG05-87ER40319 and DE-FC05-85ER2500
and by The Florida High Technology and Industry Council.} 
\bigskip
\bigskip
\cbox{Wolfhard Janke$^{1,3}$, Bernd A. Berg$^{1,2}$
            and Mohammad Katoot$^3$ }

\bigskip
\centerline{$^1$ Department of Physics}
\centerline{The Florida State University} 
\centerline{Tallahassee, Florida 32306}
\smallskip
\centerline{$^2$ Supercomputer Computations Research Institute}
\centerline{The Florida State University} 
\centerline{Tallahassee, Florida 32306}
\smallskip
\centerline{$^3$ Department of Mathematics and Physical Science}
\centerline{Embry-Riddle Aeronautical University}
\centerline{Daytona Beach, Florida 32114}
\vfil

\noindent{\bf Abstract} 
 
Using the recently proposed multicanonical ensemble, we perform Monte Carlo 
simulation for the 2d 7-state Potts model and calculate its surface free 
energy density (surface tension) to be $2 f^s = 0.0241 \pm 0.0010$. This is 
an order of magnitude smaller than other estimates in the recent literature. 
Relying on existing Monte Carlo data, we also give a preliminary estimate 
for the surface tension of 4d SU(3) lattice gauge theory with $L_t=2$. 
 \eject

\heading{Introduction} 

First-order phase transitions play a major role in many physical systems.
Many phase transitions observed in nature are first order. Examples are
vapor-liquid and liquid-solid transitions, the SU(3) deconfinement 
transition and other various stages of our universe's evolution. In a
first-order phase transition two phases can coexist at the transition 
point, with a domain wall whose tension is finite separating the two
phases. Let us denote by $f_1$ and $f_2$ the free energies per unit
volume of the two phases and by $F$ the free energy of the whole system,
the difference
$$ 
F^s\ =\ F - (V_1 f_1 + V_2 f_2)\ >\ 0   \eqno(1)
$$
is the free energy associated with the interface. Its value, normalized to
unit area $f^s$, is the surface tension, a quantity of central importance 
for the theory of first-order phase transitions~[1].

In recent years considerable efforts were devoted to surface tension
calculations. In particular, the 2d 7-state Potts model has served
as a testing ground for methods [2,3] which were subsequently applied 
to SU(3) lattice gauge theory [4,5]. Here we rely on the method of
Binder [6] to evaluate again the surface tension for these models. Our
surface tension estimates come out much lower than those of Ref.[2-5].
We explain this by showing that their extrapolations were obtained
from temperatures for which no true two phase regions exist. Our SU(3) 
estimate relies on the $L_t=2$ data of [7]. For the Potts model we have 
performed a new high statistics Monte Carlo (MC) simulation 
with the recently proposed [8] multicanonical ensemble. For the 10-state 
Potts model the multicanonical approach had improved standard MC 
calculations in the two phase region by almost three orders of magnitude [9].
In our case of the 7-state model, we find the less dramatic improvement
of a factor of two to three. This is due to the fact that the first-order
transition is considerably weaker than that of the 10-state model.

Besides serving as a testing ground for surface tension calculations, 2d 
Potts models have also attracted recent attention through their use as a 
laboratory for  testing
finite-size scaling (FSS) concepts. Rigorous work by Borgs
{\it et. al.} [10] has greatly refined the phenomenological two gaussian 
peak model [11], and the question how large the simulated systems must be 
in order that asymptotic FSS limits are approached has been investigated by
a number of MC studies [12-15]. As a by-product of our surface tension 
investigation, we can supplement these studies with results from our $q=7$ 
data.

\heading{Potts Model} 

The 2d 7-state Potts model [16] is defined on an $L^2$ lattice by the 
partition function
$$ 
Z(\beta )\ =\ \sum_{\rm configurations} \exp \left( -\beta E\right) ,
\eqno(2) 
$$
$$ E\ =\ - \sum_{<i,j>} \delta_{q_i,q_j},~~\ 
   q_i, q_j\ =\ 1,\ldots,7 . \eqno(3) 
$$
Here $i,j$ are labels of the sites of the lattice and $<i,j>$ is the sum
over nearest neighbours. We always employ the periodic boundary condition. 
Following [6] we are going to extract the
surface tension from the probability density $P_L(E;\beta )$ of the
energy $E$. For $\beta$ sufficiently close to the transition point
$P_L(E;\beta  )$ exhibits a double peak structure. We adopt the normalization
$P_L^{\max}=1$ and we define $L$-dependent pseudo-transition points 
$\beta_L^c$ by the requirement that both maxima are of equal height:
$$ 
P^{1,\max}_L = P_L (E^{1,\max}_L) = P_L (E^{2,\max}_L) = P^{2,\max}_L 
   = 1, ~~(E_L^{1,\max} < E_L^{2,\max}) . \eqno(4) 
$$
As in this equation we suppress in the following $\beta$ in the argument 
of $P_L(E;\beta )$ when the probability density is considered at its
pseudo-transition point. Once the energy probability density $P_L(E)$ is 
known, the interfacial free energy per unit
area follows [6] from the $L\to\infty$ limit of\footnote*{To comply with
[2-5] our present definition differs by a factor of two from [9].}
$$ 
2 f_L^s\ =\ -{1\over L} \ln \left( P_L^{\min} \right) . \eqno(5) 
$$
Here $P_L^{\min}=P_L(E_L^{\min})$, where $E_L^{\min}$ is the position of
the minimum which the probability density takes in the range 
$E_L^{1,\max} \le E \le E_L^{2,\max}$. The factor 2 accounts for the fact that
for periodic boundary conditions the minimum of the probability density is
governed by at least two interfaces. 
Our empirical results of $P_L(E)$ for lattices in the range $L=20,\ldots,100$
are depicted in Figure~1 on a logarithmic scale. To enhance the MC statistics
at the minima of the probability densities, we have simulated the
multicanonical ensemble as described in [8,9]. The multicanonical 
probability densities $P'_L(E)$ of our ``raw'' data are fairly 
flat in $E$ and the results of Figure~1 are obtained by reweighting for 
the correct Boltzmann factor. For our $L=100$ lattice this procedure is 
illustrated by Figure~2. The
Table gives an overview of our data and results. The statistics is given 
in sweeps, a sweep being defined as updating each spin in the lattice
once. The other quantities given are $\beta_L^c$, $e_L^{1,\max}$,
$e_L^{\min}$, $e_L^{2,\max}$ and $f_L^s$, where we denote by
$e = E / V$ the energy density and $e_L^{1,\max}$, $e_L^{\min}$ 
and $e_L^{2,\max}$ are the densities belonging to the corresponding 
extensive quantities as introduced above.

\fight=9cm
\figins{Energy density distributions $P_L(E)$ for lattices 
of size $L=20,\ldots,100$ on a logarithmic scale. The
values of the maxima have been normalized to 1.}
\figins{Multicanonical energy distribution $P'_{100} (E)$
together with its re\-weighted distribution 
$P_{100} (E)$. Both distributions are normalized
to unit area.}

\midinsert
\centerline{\bf Table: Results of multicanonical simulations}
$$\vbox{\settabs 7 \columns
\+   &         &     &                              &                 &    
                                                                      &\cr
\+ ~$L$& Statistics &~$\beta^c_L$ & $e_L^{1,\max}$ & $e_L^{\min}$ 
                                 & $e_L^{2,\max}$ & $2 f_L^s$ &\cr 
\+ ~20& ~4,000,000    & 1.284692         & 1.066(10)            & 1.3124(24)
                                 & 1.602(38)            & 0.03645(48)     &\cr
\+ ~40& ~4,000,000    & 1.291051         & 1.127(18)            & 1.327(24)
                                 & 1.578(15)            & 0.03387(59)     &\cr
\+ ~60& ~8,000,000    & 1.292283         & 1.1525(03)           & 1.3369(87)
                                 & 1.575(02)            & 0.02997(60)     &\cr
\+ 100& 16,000,000    & 1.293089         & 1.1736(29)           & 1.3466(12)
                                 & 1.570(02)            & 0.02816(58)     &\cr
\+    & & & & & & & \cr}$$
\endinsert

For lattices of size up to $L=60$ 
we have also performed simulations with a standard heat-bath
algorithm to evaluate the improvements due to the
multicanonical method. The relative performance of the two algorithms is
best evaluated by comparing the tunneling times. As in [12,15] we define 
four times the tunneling time $4\tau^t_L$ as the average number of sweeps 
needed to get from a configuration with energy $E_L^{1,\max}$ to a 
configuration with $E_L^{2,\max}$ and back. This definition has the
advantage that tunneling time and exponential autocorrelation time in
general agree with good precision (exactly in a simple model). In Figure~3 
we display on a log--log scale the divergence in the tunneling times for 
the multicanonical algorithm (circles) and for the heat bath algorithm 
(crosses). The fit to the multicanonical data is $\tau_t = 0.082(17) \times
L^{2.65(5)}$. Incidentally, this is the same power law which was observed 
in [9] for the 10-state model. Due to the weakness of the transition, 
the exponential divergence of the canonical tunneling times is not yet 
obvious in the present case. Still the multicanonical improvement should
be appreciated as we spent most of our CPU time simulating the
$100^2$ lattice.

\figins{Tunneling times for the multicanonical and for the
standard heat bath simulation on a double log scale.}

In the forthcoming fits we rely only on our multicanonical data, as 
otherwise the small lattices would get a too large weight as compared
with our $L=100$ lattice. A major point of this paper is the analysis of 
the $f_L^s$ data given in the Table. The FSS fit
$$ 
2 f^s\ =\ 2 f^s_L + {c\over L} \eqno(6) 
$$
is depicted in Figure~4 and gives
$$ 
2 f^s\ =\ 0.0241 \pm 0.0010 , \eqno(7) 
$$
\figins{Finite-size scaling fit for the interfacial free 
energy $2 f^s$. For comparison also the data from 
the standard heat bath simulation are shown.}
with $c = 0.384 \pm 0.055$ for the constant. Our result (7) is about 
a factor of ten smaller than the estimates given in [2,3]. We like to
argue that their methods analyze 
properties of rigid domain walls which have no connection
to the statistical fluctuations of the systems at their pseudo-transition
points $\beta_L^c$. To enhance the signal for the surface free
energy, the methods [2-5] have in common that they introduce domains
corresponding to temperatures $\beta^c \pm \delta$, where in the case of
the 2d 7-state Potts model the exactly known [16] transition temperature 
$\beta^c = \ln (1+\sqrt{7} ) = 1.293562\ldots$ is used. Then, it is 
tried to extract the surface tension through carefully monitoring the 
limit $\delta\to 0$. This seems precisely why their methods fail.
The signal is lost before the relevant $\delta$ region of adiabatic 
distortions is reached. For instance, the smallest $\delta$ value on 
which the extrapolation of [2] relies is $\delta = 0.01$ on a large 
$128\times 256$ lattice, where also an unused data point with 
$\delta =0.005$ exists. The $\delta =0.005$ data point had to remain 
unused as it exhibits a cross-over to much smaller surface tension values, 
interpreted then as a finite lattice size effects. 
Our Figure~5 reveals that even the $\delta = 0.005$ value 
is still too big in the sense that $\beta^c \pm 0.005$ cannot be associated
with the phase transition region. When we reweight the canonical energy 
density distribution from our $100\times 100$ lattice (where the transition
is less sharp than on the $128\times 256$ lattice) to the values 
$\beta^c\pm 0.005$ the double peak structure disappears: for $\beta^c-0.005$ 
the system is completely in the disordered, and for $\beta^c+0.005$ it is 
completely in the ordered phase. 
\figins{Reweighting of our $P_{100}(E)$ probability density
to $\beta^c - 0.005$ (high left peak alone)
and $\beta^c + 0.005$ (high right peak alone).}

\figins{Scaling of the pseudo transition temperatures
$T^{B,\min}$ and $T^{c,max}$. The solid lines are 
the exactly known asymptotic expansions.}
\figins{Results for the Binder-parameter minimum.
The arrow shows the exact infinite volume limit.}

Finally, in this section, we summarize our FSS analysis of standard
quantities like various pseudo-transition temperatures, the Binder
cumulant, the specific heat and the latent heat. Conventionally,
pseudo-transition temperatures $T^{B,\min}$ and $T^{c,max}$ are defined by
the location of the Binder-parameter minimum and specific-heat maximum,
respectively. Figure~6 shows the FSS of these quantities, together with the
exactly known [10,13] asymptotic expansions to first order. Figure~7 gives
the FSS of the Binder-parameter minimum. The arrow shows the exactly known
[12,13] infinite volume limit. In Figure~8 we show the FSS of the maxima
and the minima of the probability distribution $P_L(E)$. Linear fits to the
multicanonical data yield $e^{1,\max} = 1.2053 \pm 0.0070$ and $e^{2,\max}
= 1.5640 \pm 0.0095$, in good agreement with the exact results [17]
$1.2013\ldots$ and $1.5546\ldots$, respectively. For the minimum we find
$e^{\min} = 1.3596 \pm 0.249$. Lastly, the FSS fit of the latent heat is
given in Figure~9. The linear fit gives $\Delta e = 0.3574 \pm 0.0087$, in
agreement with the exact result $0.3533\ldots$ 

\figins{Finite-size scaling of the extrema of the 
probability distribution $P_L(E)$. }
\figins{Finite-size scaling of the latent heat. }
Besides the surface tension, the only quantities not known exactly are 
the specific heats $c_o$, $c_d$
in the ordered and disordered phase, respectively. To calculate these
quantities, we use the FSS relation [13,15]
$$ 
c^{\max}_L\ =\ L^2 \left( {\triangle s \over 2} \right)^2 + 
\triangle c + {1\over 2} (\triangle c - \triangle s) \ln (q) + c_o
+ {\cal O} \left( {1\over L^2} \right) , \eqno(8) 
$$
where $\triangle c = c_d - c_o$ and $\triangle s = \beta^c \triangle e$
is the entropy gap over the transition point. For Potts models
$\triangle c = \beta^c \triangle s / \sqrt{q}$ [17]. Figure~10 depicts the 
FSS of the specific-heat maximum with the exactly known leading term of the 
asymptotic expansion subtracted. The fit yields then $c_o = 47.5 \pm 2.4$
for the specific heat in the ordered phase. 
\fight=8cm
\figins{Finite-size scaling of the specific-heat maximum
with the exactly known leading term of the asymptotic
expansion subtracted.}
 
\heading{SU(3) Lattice Gauge Theory} 

We consider 4d pure SU(3) lattice gauge theory defined on an $L_t L^3$,
($L_t\le L$) hypercubic lattice. To each link $l$ of the lattice an
element $U_l \in {\rm SU(3)}$ is assigned and the partition function is 
given by [18]
$$ 
Z(\beta )\ =\ \int \prod_l dU_l\ \exp \left[ {2 \over g^2} 
\sum_p {\rm Re\ Tr} U({\dot p}) \right] . \eqno(9) 
$$
Here $\sum_p$ denotes the sum over all plaquettes of the lattice. For 
each plaquette $p$, $U({\dot p})$ is the ordered product of the four link 
matrices surrounding the plaquette and $dU$ is the SU(3) Hurwitz measure.
The results of the previous section imply, of course, that we cannot have 
confidence in the SU(3) surface tension estimates [4,5]. The $L_t=2$ 
SU(3) deconfining phase transition is sufficiently strong to warrant new 
MC simulations based on the multicanonical ensemble [8,9]. So far such 
calculations have not been performed. Nevertheless we are able to give a 
rough estimate by analyzing the $L_t=2$ SU(3) data of
reference [7] with regard to the surface tension, although the quality 
of the SU(3) data is fairly limited. For comparison, in case of the 7-state 
Potts model we have about 400 tunneling events for our $100^2$ lattice,
whereas we are down to only (altogether) 10 tunneling events for the 
largest SU(3) lattice, which is $2\cdot 12^2$. Available lattice sizes are 
now $L=6,8,10$ and 12. Figure~11 depicts the appropriately reweighted
probability densities for the 
lattice averages, called $s_p$, of the normalized plaquette action 
${\rm Re\ Tr} U({\dot p})/3$ (there exist two $L=12$ data sets which we
have combined to one). In contrast to the Potts model there is 
now some ambiguity in choosing the histogram binning
size, but we have checked
that within reasonable limits the influence on the final estimates is 
mild. The interfacial free energy per unit area follows in a 
similar fashion as in section~2. Equation~(5) modifies to
$$ 
2 f^s_L\ =\ -{a^{-3} \over L_t L^2} \ln (P_L^{\min}) , \eqno(10) 
$$
where $a$ is the lattice constant. It is conventional [4,5] to give
final estimates for the dimensionless quantity $f^s/T_c^3$, where
$T_c = a^{-1} L_t^{-1}$ is the physical temperature of the SU(3)
deconfining transition. Our estimates of $f^s_L/T_c^3$ are depicted 
in Figure~12. The increase of $f^s_L$ with lattice size, as compared
to the decrease found in section~2 and other Potts model investigations 
[9,12,13,15], is remarkable. Presently, we have no theoretical understanding
about the circumstances which differentiate an approach of the asymptotic 
limit from below versus above. The FSS fit of Figure~12 gives the
asymptotic estimate
$$ 
{f^s \over T_c^3}\ =\ 0.071 \pm 0.008 . \eqno(11) 
$$
Of course, in view of the rather small lattice sizes and the somewhat
limited quality of the data, this extrapolation is kind of daring.
Nevertheless, it is presumably a reasonable starting point. With 
appropriate CPU time funding of about 1,000h Cray Y-MP (or equivalent)
supercomputer time, one may perform a multicanonical simulation of
a $2\cdot 24^3$ system to check for consistency by adding  
$f^s_{24}/T_c^3$ to Figure~12. 

\figins{SU(3) action density distributions $P_L(s_p)$ for 
lattices of size $2\cdot 6^3,\ldots,2\cdot 12^3$ 
on a logarithmic scale. The values of the maxima 
have been normalized to 1.}
\figins{Finite-size scaling fit for the SU(3) 
interfacial free energy $f^s/T_c^3$.}

Comparison with Ref.[4.5] is somewhat subtle, as changing normalizations 
have to be traced. We believe that their final definitions match with 
ours and we find our result to be about a factor of two smaller than 
theirs. The discrepancy is rather mild when compared to the one of 
section~2. Considering the relative weakness of our estimate (11), one
may even question whether there is a significant difference at all.
However, the investigation of section~2 suggests that there is no reason 
to expect our estimates and those of the methods of [2-5] to converge
to the same numbers. 

\heading{Conclusions} 

We have carried out a fairly detailed FSS investigation for the 2d
7-state Potts model. As far as comparisons with exact analytical
results are possible, we find our estimates well consistent. The
surface tension is not known exactly. We give an estimate (7) which 
differs considerably from results reported in previous literature [2,3]. 
It may look kind of surprising that no problems were noted before, however, 
the methods of [2-5] seem in some sense  to be self-consistently wrong. 
They rely on simulations
far away (in the sense explained in section 2) from the relevant 
pseudo-transition $\beta$-values. The orders of magnitude of these 
deviations from the pseudo-transition points are dictated by the 
need to increase the signal, and in this region of rigid domains the
illusion of a consistent approach is, indeed, suggested by the data. 
For SU(3) our surface tension estimate (11) should be considered as
a first attempt to address the problem, and multicanonical simulations 
[8,9] on lattices of size up to at least $2\cdot 24^3$ are suggested.

\eject
\noindent{\bf Acknowledgements} 
\smallskip
We thank Thomas Neuhaus for contributions at the early stage of the Potts
model simulation. The Monte Carlo data were produced on the SCRI cluster 
of fast RISC workstations and at the Embry-Riddle Aeronautical University.

\bigskip
\noindent{\bf References}  

\item{1)} J.D. Gunton, M.S. Miguel and P.S. Sahni, in {Phase Transitions 
and Critical Phenomena}, Vol. 8, C. Domb and J.L. Lebowitz (Eds.),
Academic Press 1983.

\item{2)} J. Potvin and C. Rebbi, Phys. Rev. Lett. 62 (1989) 3062. 

\item{3)} K. Kajantje, L. K\"arkk\"ainen and K. Rummukainen, Phys.
          Lett. B223 (1989) 213.

\item{4)} K. Kajantje, L. K\"arkk\"ainen and K. Rummukainen, Nucl. Phys.
          B333 (1990) 100; B357 (1991) 693.

\item{5)} S. Huang, J. Potvin, C. Rebbi and S. Sanielevici,
          Phys. Rev. D42 (1990) 2864; D43 (1991) 2056 (E).

\item{6)} K. Binder, Phys. Rev. A25 (1982) 1699; 
          K. Binder, Z. Phys. B43 (1981) 119.

\item{7)} N.A. Alves, B.A. Berg and S. Sanielevici, preprint, SCRI-91-93,
          to be published in Nucl. Phys. B.

\item{8)} B.A. Berg and T. Neuhaus, Phys. Lett. B267 (1991) 249.

\item{9)} B.A. Berg and T. Neuhaus, Phys. Rev. Lett. 68 (1992) 9.

\item{10)} C. Borgs and R. Koteck\'y, J. Stat. Phys. 61 (1990) 79;
           C. Borgs, R. Koteck\'y and S.~Miracle-Sole, J. Stat. 
           Phys. 62 (1991) 529.

\item{11)} K. Binder and D.P. Landau, Phys. Rev. B30 (1984) 1477;
           K. Binder, M.S. Challa and D.P. Landau, Phys. Rev. 
           B34 (1986) 1841.

\item{12)} A. Billoire, S. Gupta, A. Irb\"ack, R. Lacaze, A.
           Morel and B. Petersson, Nucl. Phys. B358 (1991) 231;
           A. Billoire, R. Lacaze and A. Morel, preprint, 
           SPhT-91/122, to be published in Nucl. Phys. B.

\item{13)} J. Lee and J.M. Kosterlitz, Phys. Rev. B43 (1991) 3265.

\item{14)} C. Borgs and W. Janke, preprint, FUB-HEP 6/91, to be
           published in Phys. Rev. Lett.

\item{15)} W. Janke, preprints, HLRZ, J\"ulich (1992).

\item{16)} R.B. Potts, Proc. Cambridge Philos. Soc. 48 (1952) 106;
           F.Y. Wu, Rev. Mod. Phys. 54 (1982) 235.

\item{17)} R.J. Baxter, J. Phys. C6 (1973) L445. 

\item{18)} K. Wilson, Phys. Rev. D10 (1974) 2445.

\bye
%%%%
%  cover page
%%%%%
\input preprint92.format
%  file to do disclaimers and cover pages  ONLY..   the commands used
%   in this file are located at the bottom of this file.  Comment out
%   those commands you DO NOT use. 
%%%%

\covertitle{MONTE CARLO CALCULATION OF THE SURFACE FREE ENERGY
FOR THE 2D 7-STATE POTTS MODEL, AND AN ESTIMATE FOR 4D SU(3) GAUGE THEORY}
\coverauthor{Wolfhard~Janke, Bernd~A.~Berg, and~Mohammad~Katoot}
\coverpreprint{40}
\coverdate{February 1992}

\makecover
\bye